\documentclass[final,3p,times,twocolumn]{elsarticle}

\usepackage{graphicx}

\usepackage{amssymb}

\newcommand{\leff}{$\mathcal{L}_{eff}~$}
\newcommand{\Qy}{$\mathcal{Q}_y~$}

\begin{document}

\begin{frontmatter}



\title{The scintillation and ionization yield of liquid xenon for nuclear recoils\tnoteref{t1}}


\author[brow]{P.~Sorensen\corref{cor1}} \ead{pfs@het.brown.edu} 
\author[yale]{A.~Manzur} 
\author[prin]{C.E.~Dahl}
\author[zuri,flor]{J.~Angle} 
\author[colu]{E.~Aprile}
\author[lngs]{F.~Arneodo} 
\author[zuri]{L.~Baudis} 
\author[llnl]{A.~Bernstein}
\author[case]{A.~Bolozdynya}
\author[coim]{L.C.C.~Coelho}
\author[brow]{L.~DeViveiros}
\author[zuri,lngs]{A.D.~Ferella}  
\author[coim]{L.M.P.~Fernandes}
\author[brow]{S.~Fiorucci}
\author[brow]{R.J.~Gaitskell}
\author[colu]{K.L.~Giboni} 
\author[rice]{R.~Gomez}
\author[yale]{R.~Hasty}
\author[yale]{L.~Kastens}
\author[prin]{J.~Kwong}
\author[coim]{J.A.M.~Lopes}
\author[llnl]{N.~Madden}
\author[zuri,flor]{A.~Manalaysay}
\author[yale]{D.N.~McKinsey}
\author[colu]{M.E.~Monzani}
\author[yale]{K.~Ni}
\author[rice]{U.~Oberlack}
\author[aach]{J.~Orboeck}
\author[yale]{G.~Plante}
\author[colu]{R.~Santorelli}
\author[coim]{J.M.F.~dos~Santos}
\author[rice]{P.~Shagin}
\author[case]{T.~Shutt}
\author[aach]{S.~Schulte}
\author[llnl]{C.~Winant} 
\author[colu]{M.~Yamashita} 

\address[brow]{Department of Physics, Brown University, Providence, RI 02912, USA}
\address[case]{Department of Physics, Case Western Reserve University, Cleveland, OH 44106, USA}
\address[colu]{Department of Physics, Columbia University, New York, NY 10027, USA}

\address[lngs]{Gran Sasso National Laboratory, Assergi, LÕ'Aquila, 67010, Italy}
\address[llnl]{Lawrence Livermore National Laboratory, 7000 East Ave., Livermore, CA 94550, USA}
\address[prin]{Department of Physics, Princeton University, Princeton, NJ 08540, USA}
\address[rice]{Department of Physics, Rice University, Houston, TX 77251, USA}
\address[aach]{Department of Physics, RWTH Aachen University, Aachen, 52074, Germany}
\address[yale]{Department of Physics, Yale University, New Haven, CT 06511, USA}
\address[flor]{Department of Physics, University of Florida, Gainesville, FL 32611, USA}  
\address[coim]{Department of Physics, University of Coimbra, R. Larga, 3004-516, Coimbra, Portugal}
\address[zuri]{Physics Institute, University of Z\"urich, Winterthurerstrasse 190, CH-8057,  Z\"urich, Switzerland}
 
\cortext[cor1]{Corresponding author}
\tnotetext[t1]{XENON10 Collaboration}

\begin{abstract}
XENON10 is an experiment designed to directly detect particle dark matter.  It is a dual phase (liquid/gas) xenon time-projection chamber with 3D position imaging.  Particle interactions generate a primary scintillation signal ($S1$) and ionization signal ($S2$), which are both functions of the deposited recoil energy and the incident particle type.  We present a new precision measurement of the relative scintillation yield $\mathcal{L}_{eff}$ and the absolute ionization yield $\mathcal{Q}_y$, for nuclear recoils in xenon.  A dark matter particle is expected to deposit energy by scattering from a xenon nucleus.  Knowledge of \leff is therefore crucial for establishing the energy threshold of the experiment; this in turn determines the sensitivity to particle dark matter.  Our \leff measurement is in agreement with recent theoretical predictions above 15~keV nuclear recoil energy, and the energy threshold of the measurement is $\sim4$~keV.  A knowledge of the ionization yield \Qy is necessary to establish the trigger threshold of the experiment.  The ionization yield \Qy is measured in two ways, both in agreement with previous measurements and with a factor of 10 lower energy threshold.
\end{abstract}


\end{frontmatter}



\section{Introduction}
\subsection{Expected signal in XENON10}
There is abundant evidence for a significant cold dark matter (CDM) component in the universe \cite{freedman2003,bullet_cluster,dm_ring}, and perhaps the best-motivated candidate is the lightest neutralino from super-symmetric (SUSY) extensions to the Standard Model \cite{jungman1996}.  A neutralino is expected to be non-relativistic and stable, and is more generally classified as a weakly interacting massive particle (WIMP).  The open question of the expected mass and cross-section of WIMPs is being addressed by numerous direct and indirect detection experiments \cite{gaitskell2004,gabriel05,laura06}, including XENON10.

The XENON10 detector is a liquid xenon time-projection chamber.  It is designed to directly detect galactic WIMPs which scatter elastically from xenon nuclei.  With velocities of order 10$^{-3}$c, the recoil energy spectra WIMPs with a mass 100~GeV/c$^2$ incident on xenon is predicted to be a featureless exponential falling one decade every 30~keV nuclear recoil energy (keVr).  A particle interaction in liquid xenon creates both excited and ionized xenon atoms \cite{1978Kubota}, which react with the surrounding xenon atoms to form excimers.  The excimers relax on a scale of $10^{-8}$~s with the release of scintillation photons.  This prompt scintillation light is detected and referred to as the $S1$ signal.  

An external electric field ($E_d=0.73$~kV/cm) across the liquid xenon target causes a portion of the ionized electrons to be drifted away from an interaction site.  The electrons are extracted into the gas phase by a stronger electric field ($\sim10$~kV/cm) and accelerated through a few mm of xenon gas, creating a secondary scintillation signal.  This scintillation light is proportional to the number of ionized electrons and is referred to as $S2$.  The amplification during proportional scintillation makes the recoil energy threshold for $S2$ lower than the threshold for $S1$.  XENON10 discriminates between electron recoil background and the expected nuclear recoil WIMP signal via the distinct ratio of ionization ($S2$) to scintillation ($S1$) for each type of interaction.

\subsection{Importance of these measurements} \label{importance}
The energy threshold of XENON10 is determined by its total light collection efficiency for primary scintillation photons ($S1$), and by the effective scintillation yield of nuclear recoils ($\mathcal{L}_{eff}$).  Because of the exponential slope of the expected signal, the detector energy threshold bears significantly on the ultimate sensitivity of XENON10.  The sensitivity of XENON10 to spin-independent interactions \cite{2008XENON10SI} and spin-dependent interactions \cite{2008XENON10SD} are reported in separate letters, based on a constant \leff=~0.19.  Several groups have measured \leff using tagged neutron scattering, with a range of results \cite{aprile2005, chepel2006}.  

In Sec. \ref{leff_method} we present an alternative method to measure $\mathcal{L}_{eff}$. We clearly establish the energy dependence of \leff in the range $4-100$~keVr.  The uncertainty is substantially reduced compared with previous measurements.  In Sec. \ref{subsecEvent} $-$ Sec. \ref{S1_res_systematics} we make a careful assessment of possible systematic and statistical uncertainties affecting our measurement.  The effect on the dark matter sensitivity of XENON10 is discussed in Sec. \ref{summary_sec}.

In Sec. \ref{e_yield} we report a measurement of the absolute ionization yield ($\mathcal{Q}_y$) of nuclear recoils in liquid xenon.  Our results are in agreement with previous measurements, above 25~keVr \cite{aprile2006}.  To our knowledge, this is the first measurement of \Qy below 25~keVr.  In Sec. \ref{msmethod} we also present a new method to determine the absolute ionization yield.  This method provides a cross-check on our measurement of $\mathcal{L}_{eff}$.

\section{Experimental Apparatus} \label{exp}
\subsection{XENON10 detector and neutron calibration}
\begin{figure}[h]
\includegraphics[width=0.48\textwidth]{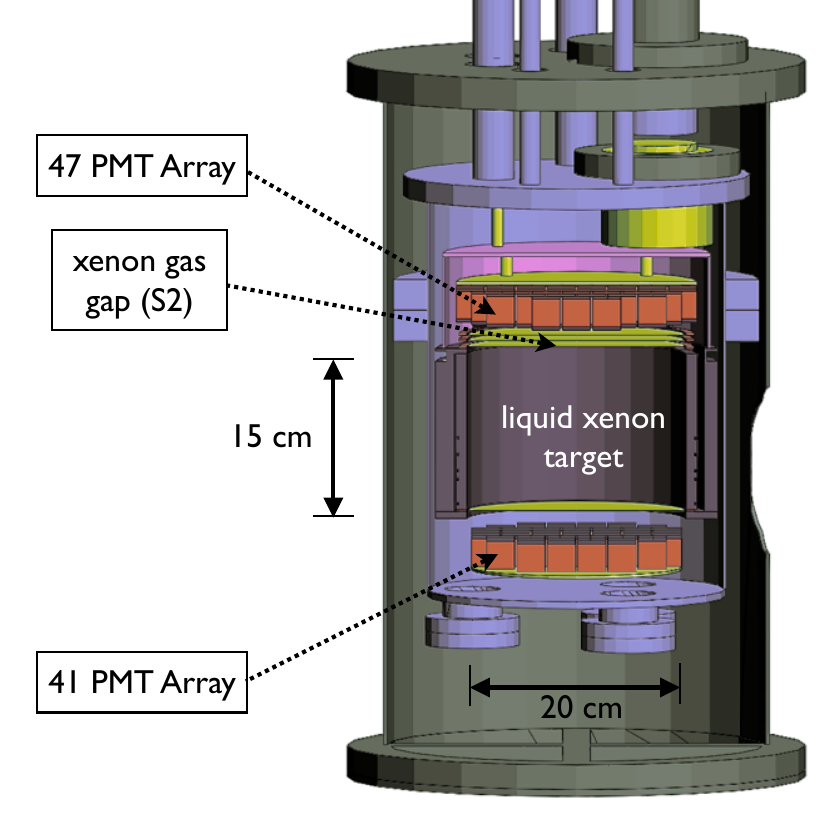}
\caption{A side view cut-away schematic of the XENON10 detector as rendered by the GEANT4 simulation.  The liquid xenon target is 15~cm in height with a 20~cm diameter.   Ancillary systems, cabling, shielding etc. are omitted for clarity.  The detector was completely enclosed by 20~cm Pb outside of 20~cm polyethylene shielding.}
\label{fig1}
\end{figure}

XENON10 is a position-sensitive time-projection chamber.  Two arrays of UV-sensitive photomultiplier tubes (PMT) detect the $S1$ and $S2$ signals.  The XENON10 instrument, including design, energy calibration and position-dependent corrections, is described in detail in \cite{XENON10Instr}.  The performance of the 3D position reconstruction is described in \cite{2006Ni}.  The XENON10 detector is shown schematically in Fig. \ref{fig1}.  

The XENON10 detector was exposed for 12 hours, with a live fraction 0.92, to a 3.7~MBq $\pm15\%$ AmBe source emitting $220$~n/s.  The neutron rate is based on a yield of $6\times10^{-5}$~n/Bq \cite{PDG}.  The exposure occurred in low-background conditions at Laboratori Nazionali del Gran Sasso.  With $3100$ meters water equivalent rock overburden, the cosmic muon flux is reduced by about $10^6$ compared with surface conditions \cite{2006_Mei}.  The instrument was shielded by 20~cm Pb outside of 20~cm high density polyethylene.  The shield completely enclosed the detector.  It reduced the cavern $\gamma$ flux by more than $10^5$, and the cavern neutron flux by about $10^{2.5}$ \cite{XENON10Instr}.  

\subsection{Neutron source and Monte Carlo simulation}
The AmBe source was attached to a steel rod and inserted through a 7~mm diameter hole in the shield. It was positioned next to the detector cryostat, behind an additional 5~cm of Pb shielding.  The active target of XENON10 had a 10~cm radius and 15~cm height, with a xenon mass of 13.7~kg.  The analysis presented here uses only nuclear recoils which occurred in a 5.4~kg fiducial target, with 8~cm radius and 9.3~cm height.  This is the same fiducial target used for the blind analysis of the WIMP-search data, as reported in \cite{2008XENON10SI}, surrounded by a minimum of 2~cm of self-shielding xenon.  

Initial neutron energies from the AmBe source ranged from below 0.1 MeV to 11 MeV, with a mean at 4.3 MeV \cite{marsh1995}.  The neutron energy spectrum is known with an accuracy of $\pm3\%$ (per 0.1~MeV bin) for a source of this strength \cite{thomas_pc}.  A detailed Monte Carlo simulation of the nuclear recoil spectrum in the xenon target was found to be insensitive to variations on this scale.  Despite the features in the AmBe source energy spectrum \cite{marsh1995}, the spectrum of neutron energies as they enter the 5.4~kg fiducial target is a featureless exponential, falling 1 decade in 3.5~MeV.

The source also emitted 148~$\gamma$/s at 4.4~MeV from the de-excitation of the $^{12}$C final state.  A 0.06~MeV $\gamma$ with a branching ratio of 36\% was not relevant due to the 5~cm of internal Pb shielding.  The Monte Carlo simulation predicted a flat rate of single-scatter $\gamma$ events in the fiducial target with E~$<100$~keV electron recoil equivalent energy (keVee).  The predicted rate was reduced by $\times40$ by the 5~cm of internal Pb shielding, and by an additional $\times3$ due to the 2~cm of self-shielding xenon.  Prior to $S2/S1$ discrimination, the single scatter $\gamma$ rate in the energy range E~$<100$~keVee was measured to be $<2$~cts/0.25~keVee over the full exposure.  In contrast, the single scatter elastic nuclear recoil rate was higher by as much as $\times400$, as shown in Fig. \ref{fig2}.  Inelastic nuclear recoils on xenon cause additional prominent $\gamma$ lines at 40 and 80~keVee.  These events were easily avoided by the $S2/S1$ discrimination, as shown in Fig. \ref{fig4}.

\subsection{Data acquisition and analysis} \label{daqaa}
Event data were recorded at a rate of 6.5~Hz during the 12~hour exposure.  An $S2$-sensitive trigger was obtained from the shaped ($\tau=1~\mu$s) sum of the 30 central (of 47) top-array PMTs.  The trigger was verified with a known voltage input to have full efficiency for an $S2$ corresponding to 4 electrons, and about 75\% acceptance for 3 electrons.  Analysis was restricted to $S2$ pulses corresponding to at least 8 electrons, which is more than $3\sigma$ below the typical $S2$ pulse at $S1=3$~photo-electrons.  This is shown in Fig. \ref{fig4}.  The $S1$ signal was found by look-back after events were digitized.  The $S1$ peak-finding efficiency depends on the size of individual photo-electron pulses from the PMTs, and the requirement of a minimum of 0.35~photo-electrons/PMT in 2 or more PMTs ($n\geq2$ coincidence) in a $0.3~\mu$s time window. The $S1$ peak-finding efficiency was predicted to be better than 0.985 at 1~keVee.  This is discussed further in Sec. \ref{S1_systematics}, and is indicated by the dotted line in Fig. \ref{fig5}.  A high energy veto was set to avoid digitizing $\gamma$ scatters with energies above $\sim 150$ keVee.  The veto was about $\times1.5$ above the high-energy tail of the elastic nuclear recoil spectrum.   

\section{Scintillation Yield} \label{scint_yield}
\subsection{Measurement of \leff} \label{leff_method}
The nuclear recoil energy $E_{nr}$ (in units of keVr) was determined from the measured $S1$ response (in units of keVee) according to 
\begin{eqnarray} \label{leff_def}
E_{nr} = \frac{S1}{L_y \cdot \mathcal{L}_{eff}} \cdot \frac{S_e}{S_n}
\end{eqnarray}
where \leff is the scintillation yield of xenon for nuclear recoils, relative to the zero-field scintillation yield for electron recoils at 122~keVee.  The use of this standard reference energy avoids any systematic error from the non-linear scintillation yield of xenon for electron recoils.  After position-dependent corrections, XENON10 measured a volume-averaged light yield $L_y~=~3.0~\pm~0.1$~(syst)~$\pm~0.1$~(stat)~photo-electons/keVee for 122~keVee $\gamma$ events \cite{XENON10Instr}.  Scintillation quenching by the external electric field $E_d=0.73$~kV/cm was previously measured to be $S_e=0.54\pm0.01$ for electron recoils and $S_n=0.93\pm0.04$ for nuclear recoils \cite{aprile2005}.    
\begin{figure}[h]
\includegraphics[width=0.48\textwidth]{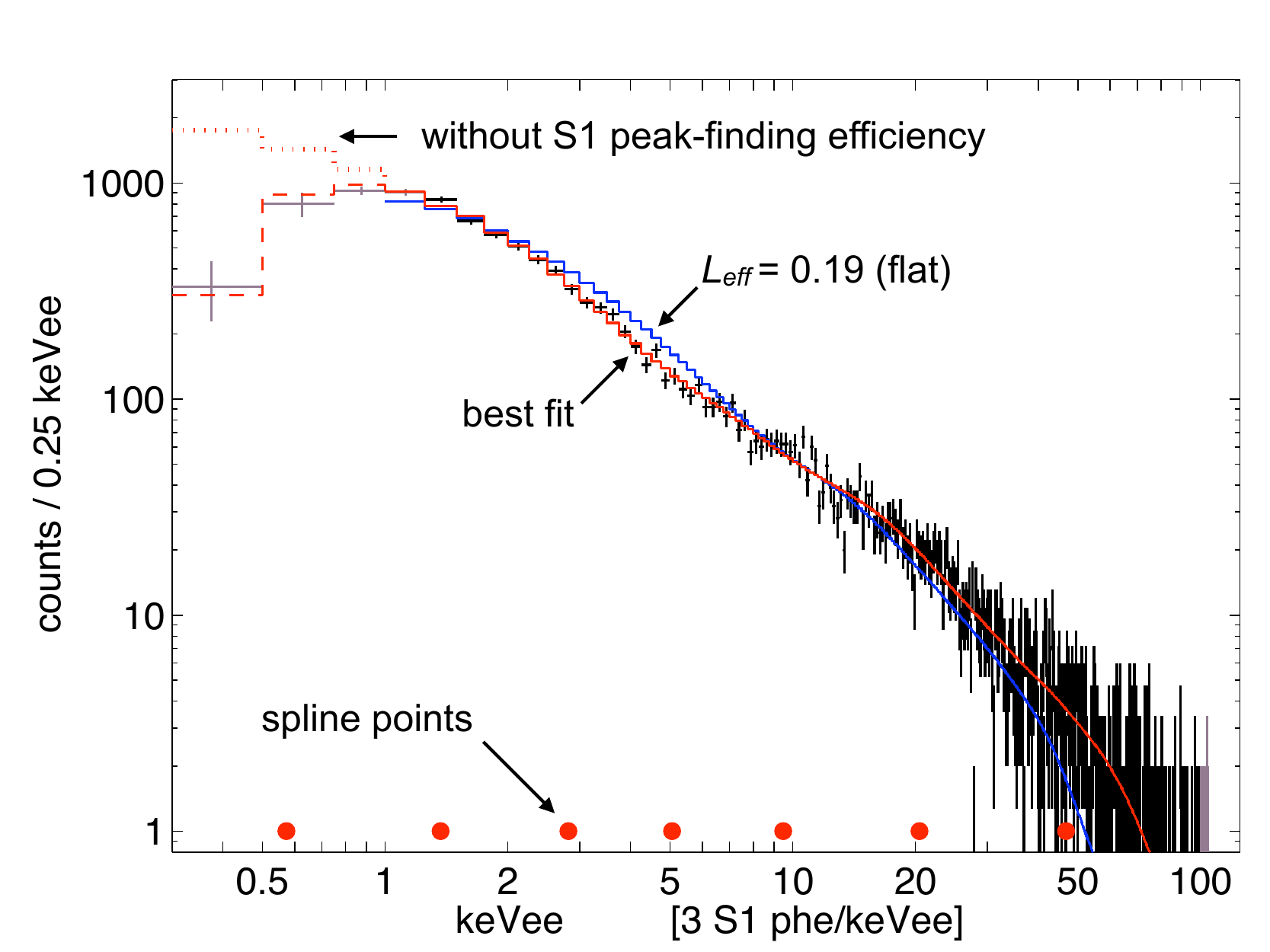}
\caption{The primary scintillation spectrum (with $1\sigma$ errors) of single scatter nuclear recoils from 11~hrs live exposure to an AmBe neutron source, and the result from a detailed Monte Carlo of the experiment (red line, labeled ``best fit''), considering the energy dependence of \leff shown in Fig. \ref{fig3}.  The $S1$ peak-finding efficiency for the bin centered on 1.125~keVee (3.4~photo-electrons) is $>0.99$.  Bins $<1$~keVee were not used to obtain the fit, and are shown in grey (data) or dashed/dotted red (Monte Carlo).  The translation from keVr to keVee is shown for the 7 \leff spline points from Fig. \ref{fig3} (red circles along axis).  Also shown are the Monte Carlo spectra without the effect of the $S1$ peak-finding efficiency (red dotted), and for a constant \leff=~0.19 (solid blue line).}
\label{fig2}
\end{figure}

\begin{figure}[h]
\includegraphics[width=0.48\textwidth]{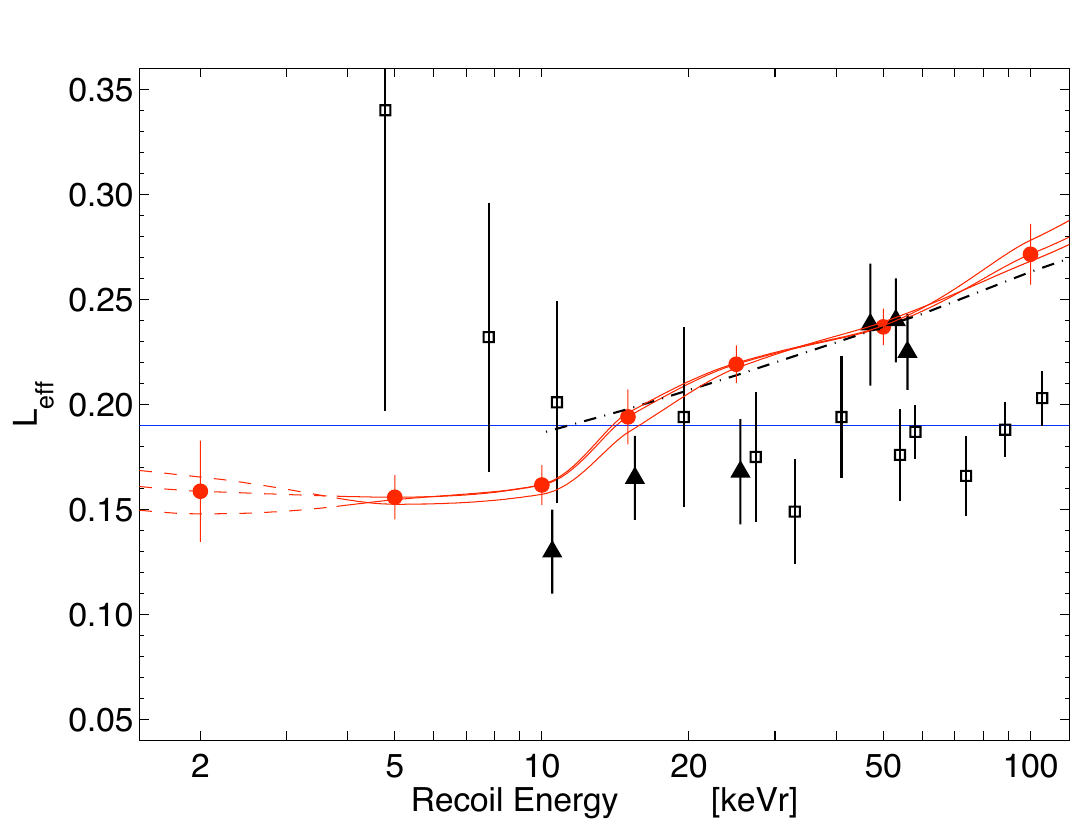}
\caption{The energy dependence of \leff at $E_d=0.73$~kV/cm (red curves) obtained from the measurement described in Sec. \ref{leff_method}, and the spline points (red circles) with $1\sigma$ (statistical) errors.  The three curves indicate the range of systematic uncertainty.  The uncertainties are summarized in Table \ref{systm}.  Also shown are data from \cite{aprile2005} (triangles), \cite{chepel2006} (squares) and the theoretical prediction of \cite{2005_Hitachi} above 10~keVr (dash-dot).  \leff is shown dashed below $\sim3.5$~keVr (1~keVee), to indicate the energy regime not used to obtain the best fit.  A constant \leff=~0.19 is also shown (solid blue line).}
\label{fig3}
\end{figure}

The Monte Carlo recoil energy spectrum was converted from keVr to keVee, via Eq. \ref{leff_def}, and convolved with the measured $S1$ energy resolution.  \leff was then varied to give the best agreement between the measured and Monte Carlo energy spectra, using the binned maximum likelihood method \cite{pdg_stats_2004}.  The primary scintillation spectrum of single-scatter nuclear recoils is shown in Fig. \ref{fig2} (black line, with $1\sigma$ error bars), along with the best-fit Monte Carlo spectrum (red line).  The fit has $\chi^2=401$ with $d.o.f.=385$, and a $p$-value of 0.28.  For comparison, the best-fit spectrum assuming a constant \leff=~0.19 is also shown (blue line,  $\chi^2=549$, $p<10^{-6}$).  To avoid additional systematic uncertainty, bins for which the $S1$ peak-finding efficiency was predicted to be $<0.99$ were not included in the maximum likelihood fit, and are indicated in grey.  The Monte Carlo spectrum for these bins is shown dashed.  The same spectrum without the S1 peak-finding efficiency is shown dotted.

The form of \leff was modeled by a piece-wise cubic spline, which provides optimal freedom with continuous $1^{st}$ and $2^{nd}$ derivatives.  Spline points were fixed at 2, 5, 10, 15, 25, 50 and 100~keVr and unconstrained in $\mathcal{L}_{eff}$.  Spline endpoints (also unconstrained in $\mathcal{L}_{eff}$) at 0.8 and 500~keVr are not shown;  $\mbox{d}^2$\leff$/\mbox{d}(E_{nr})^2=0$ at the endpoints.  Using fewer spline points was found to over-constrain the result.  Conversely, using more spline points merely added fluctuations (within the statistical error).  If the keVr values of the spline points were translated in either direction, the result shifted by less than $\pm0.005$.  This effect is included in the statistical uncertainty as shown in Table \ref{systm} and Fig. \ref{fig3}.  

The energy dependence of \leff is shown in Fig. \ref{fig3} (red lines), along with the spline points (red circles).  Also shown are data from \cite{aprile2005} (triangles), \cite{chepel2006} (squares) and the theoretical prediction of \cite{2005_Hitachi} above 10~keVr (dash-dot line).  The keVee equivalent value of each spline point is indicated in Fig. \ref{fig2} (red circles, along the axis).  The $1\sigma$ (stat) error bars on the best-fit \leff spline points were estimated by a Monte Carlo method that is standard for multi-parameter fits:  the bin counts of the best-fit recoil spectrum shown in Fig. \ref{fig2} were allowed to Poisson-fluctuate.  The resulting spectrum was then treated as input data, and fit with the same method described above.  About $10^4$ such ``experiments'' were performed, and the resulting \leff values at each of the spline points were found to be normally distributed.
 \begin{figure}[h]
\includegraphics[width=0.48\textwidth]{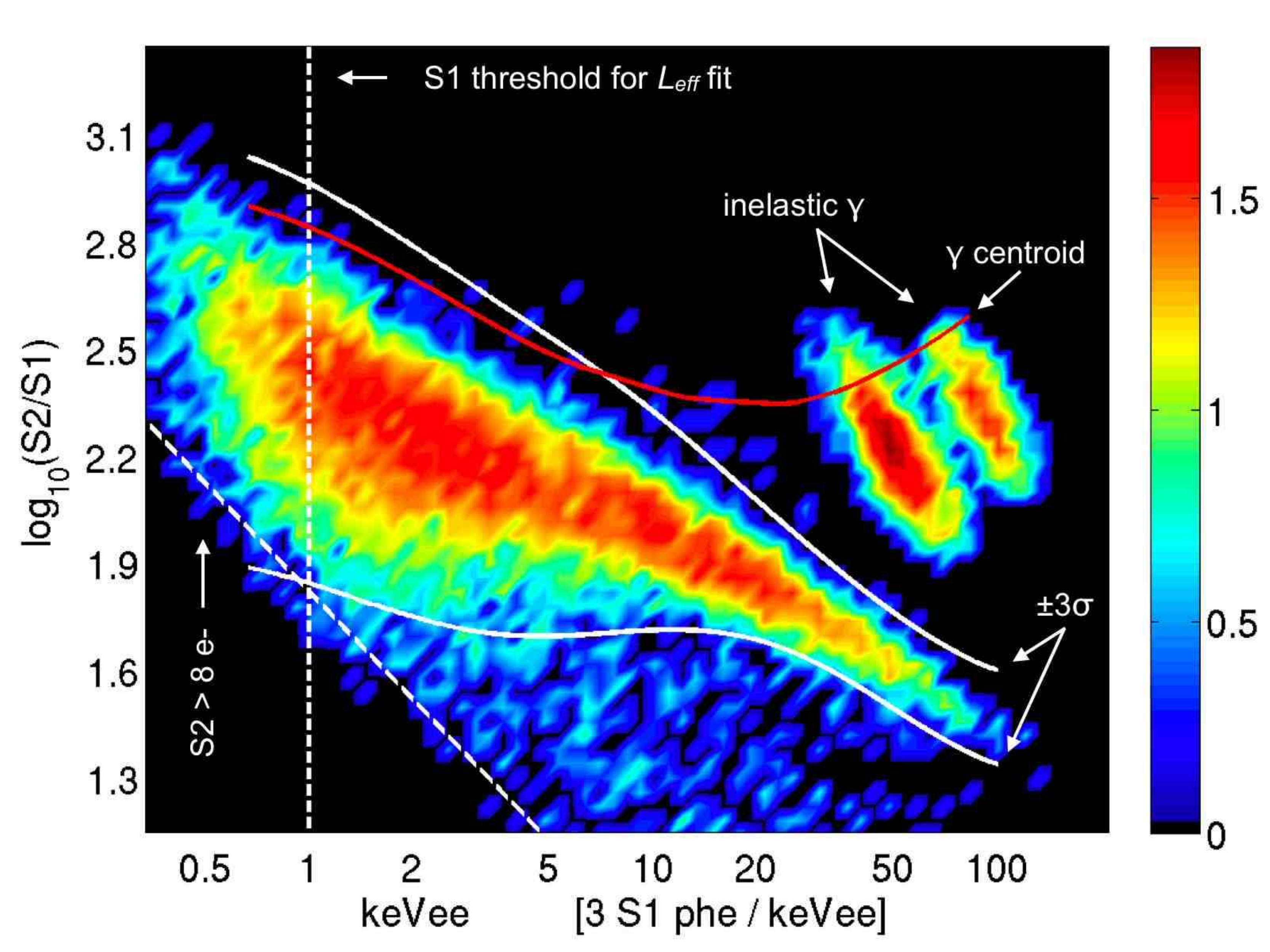}
\caption{The discrimination parameter log$_{10}(S2/S1)$ for nuclear recoils in xenon.  The color scale ($z$ axis) indicates log$_{10}$(counts). The parameter is Gaussian except for a small ($<2\%$) secondary population below the $-3\sigma$ contour.  The analysis thresholds for scintillation light ($S1$) and charge ($S2$) are indicated by dashed lines.  The centroid of the distribution for $\gamma$-induced electron recoils (red line) is expected to lie slightly above the center of the 40 and 80~keVee inelastic populations \cite{2008_Dahl}.  $\gamma$ events with $E>80$~keVee are not shown. }
\label{fig4}
\end{figure}
The overall event rate normalization between the Monte Carlo and data was treated as a free parameter in the \leff minimization.  Allowing for the DAQ dead time of 8\% and the event acceptance described in Sec. \ref{subsecEvent}, the Monte Carlo predicted an absolute single scatter neutron rate consistent with the uncertainty on the AmBe source strength.  

\subsection{Event Acceptance} \label{subsecEvent}
Event records of 156~$\mu$s were digitized with a 105~MHz ADC.  Valid events were required to consist of a single $S1$ pulse followed by a single $S2$ pulse.  The pulses are easily distinguishable by their width and shape:  $S1$ decays with $\tau=27$~ns \cite{1999_Doke}, while $S2$ is roughly Gaussian with FWHM $\sim0.8~\mu$s.   There was no distinguishable $S1$ pulse for 7\% of events inside an 8~cm radius (the $z$-coordinate is indeterminate without the $S1$).  A further 4\% of events had an $S1$ candidate pulse with all its signal concentrated in a single PMT.  Such events fail the $n\ge2$ coincidence requirement.  Both of these observations were expected due to Poisson fluctuations in the small number of photo-electrons resulting from low-energy nuclear recoils.  In particular, the number of events considered as having no $S1$ was predicted to better than 15\% below 1~keVee.  This can be seen in Fig. \ref{fig2} as the difference between the best-fit Monte Carlo (dashed), and the same spectrum without the $S1$ peak-finding efficiency applied (dotted).

Basic quality cuts placed loose constraints on acceptable $S1$ and $S2$ pulse widths, and the distribution of scintillation light between top and bottom PMT arrays.  Additionally, the first and last 50 samples of each event were required to be featureless.  These cuts had a combined acceptance $>99\%$ for genuine scatters in the fiducial target, as previously reported for the WIMP-search results \cite{2008XENON10SI}.  A $\pm3\sigma$ cut on the discrimination parameter $y=$~log$_{10}(S2/S1)$ was applied to select the elastic nuclear recoils, as indicated in Fig. \ref{fig4}.  This cut excluded a small ($<2\%$) non-Gaussian population of events below the $-3\sigma$ band in Fig. \ref{fig4}.  
\begin{figure}[h]
\includegraphics[width=0.48\textwidth]{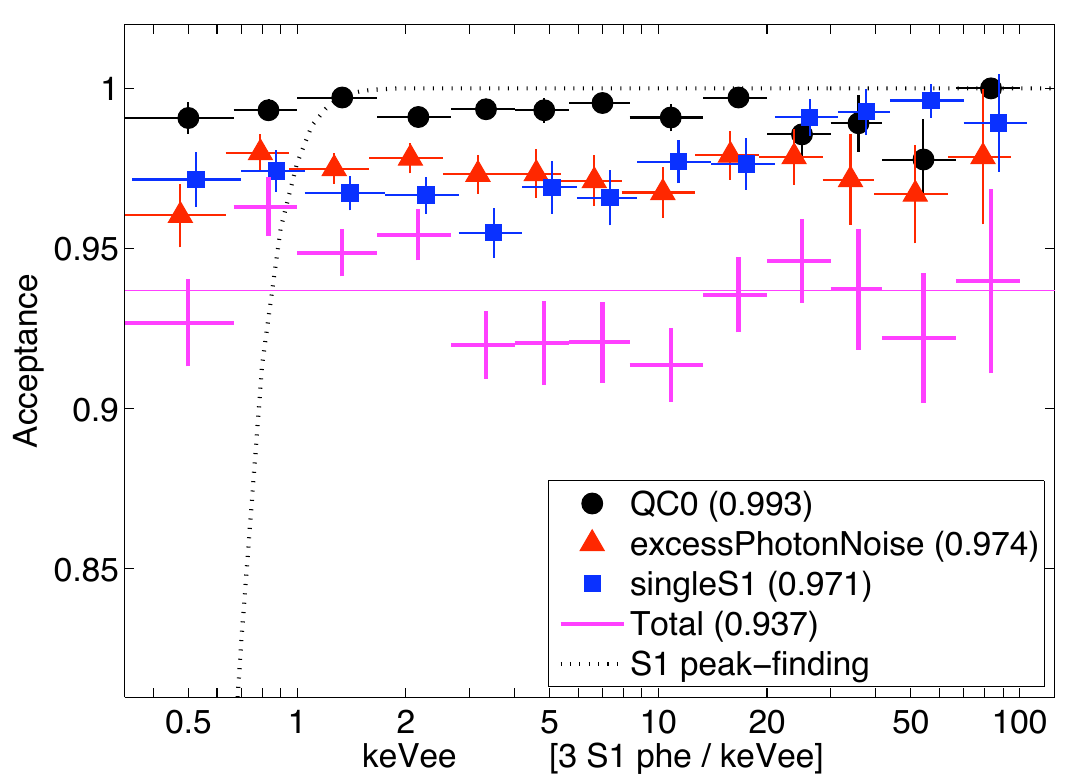}
\caption{ Event acceptance for single-scatter nuclear recoils in the fiducial target.  Average acceptance $0.5-100$~keVee is indicated in the legend.  The cuts are described in Sec. \ref{subsecEvent}.  The $S1$ peak-finding efficiency is discussed in Sec. \ref{S1_systematics}.}
\label{fig5}
\end{figure}
Two other quality cuts were applied to the data.  The first (``excess Photon Noise'') required that the proportion of signal concentrated in the primary $S2$ pulse be $>0.85$;  pulses failing this cut tend to have an excess of single-electron $S2$ events (or other spurious photon noise), which can complicate the determination of pulse parameters.  The second (``single S1'') required that only 1 possible $S1$-like pulse be found before the $S2$ pulse;  a secondary $S1$ candidate was required to have at least $\times0.25$ as many photo-electrons as the primary.  The rate of $S1$-only pulses in the absence of the AmBe source was $>100$~Hz, with an exponentially falling energy spectrum.  The spectrum of these random coincidence events leads to the mild increase in acceptance of the ``single $S1$'' cut with energy, as shown in Fig. \ref{fig5}.  Below about 3~keVee the acceptance rises again due to the $\times0.25$~photo-electron threshold for declaring multiple $S1$ candidates.   Each of these cuts had an average acceptance $>0.97$ for elastic nuclear recoils in the fiducial target.  The cuts are not completely orthogonal, and the combined average acceptance for nuclear recoils $-$ including basic quality cuts $-$ was found to be 0.937.  The measured \leff was not significantly changed if any or all of these cuts were removed, however, the quality of the fit was worse by 10-20\%.

Single-electron $S2$ pulses were found in about 50\% of elastic nuclear recoil events, not necessarily correlated with the event.  About 4\% of events also exhibited a small secondary $S2<200$~photo-electrons.  A single scatter nuclear recoil with $S2\sim200$~photo-electrons (about 8~electrons) would most likely fall below the detector $S1$ threshold, as can be seen from Fig. \ref{fig4}.  Secondary $S2$ pulses of this size or smaller were not treated as multiple scatters.

\subsection{Signal Identification and $S1$ Acceptance} \label{S1_systematics}
\begin{figure}[h]
\includegraphics[width=0.48\textwidth]{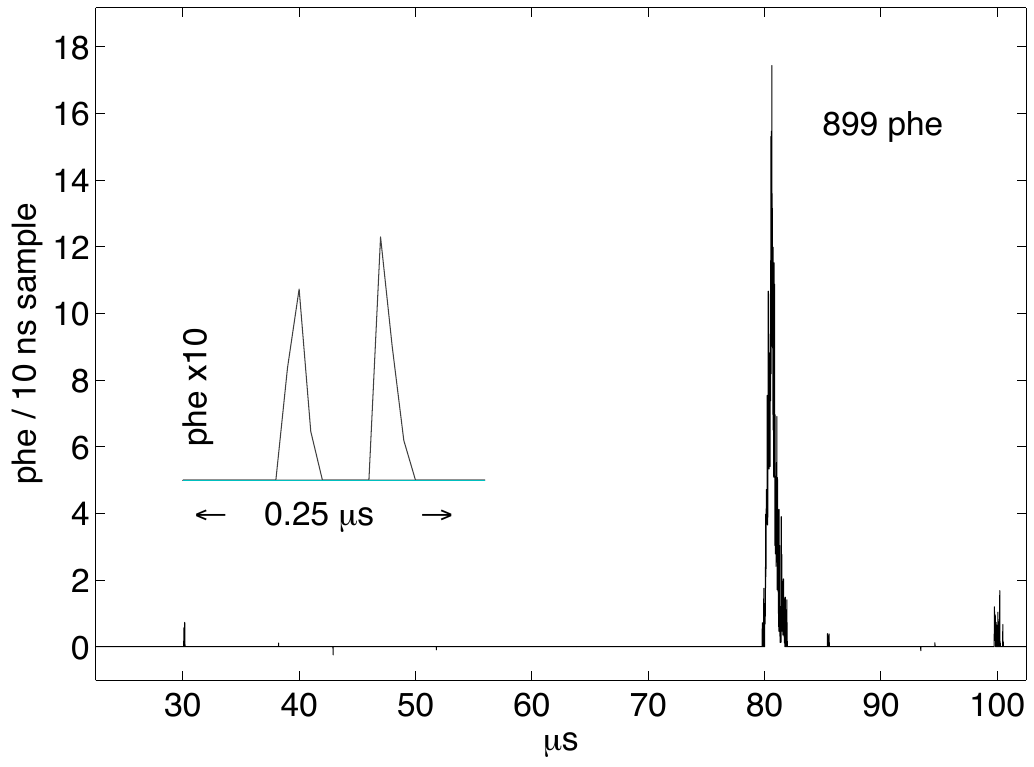}
\caption{ An elastic nuclear recoil event at 0.75~keVee.  The $S1$ signal of 2.3~photo-electrons at $30~\mu$s is also shown inset $\times10$ in $y$ and $\times100$ in $x$.  The $S2$ signal consists of 899~photo-electrons (about 37 ionization electrons).  The small pulse at $100~\mu$s is a 22~photo-electron $S2$ from a single ionization electron.  Baseline fluctuations of up to $0.10\pm0.03$~photo-electrons/sample were set to zero in the analysis.}
\label{fig6}
\end{figure}

A typical $0.75$~keVee elastic nuclear recoil event with $S1=2.3$~photo-electrons and $S2=899$~photo-electrons is shown in Fig. \ref{fig6}.  Baseline fluctuations of up to $0.10\pm0.03$~photo-electrons/sample were set to zero.  The threshold was set individually for each PMT based on RMS noise.  The waveform is shown summed across all 89 PMTs.  Coherent noise pick-up was efficiently rejected because photo-electron pulses were checked on individual PMTs.  The pulse at $100~\mu$s is a typical single-electron $S2$, consisting of $22$~photo-electrons spread out over $\sim0.8~\mu$s.
The $S1$ peak-finding efficiency shown in Fig. \ref{fig5} (dotted) was obtained by a detailed simulation of the photo-electron counting statistics, including PMT response and sampling effects, as described in Sec. \ref{daqaa}.  It predicted a flat 100\% efficiency above 2~keVee.  The low-energy Compton-scatter background observed in $\gamma$ calibration data was found to be flat in the energy range $2-20$~keVee.  

Under the assumption that this background should remain flat in the range $0.2-2$~keVee, the predicted $S1$ peak-finding efficiency and the observed roll-off in the $\gamma$ calibration data are consistent at the $1\sigma$ level.  In order to estimate the possible systematic effect of the $S1$ peak-finding efficiency on the \leff result, an $S1$ efficiency curve shifted higher (lower) by 20\% was also considered.  The effect on \leff at 2~keVr was a systematic decrease (increase) of $0.012$.  At all other spline points, the shift was at most half this amount.  This is summarized in Table \ref{systm}.  As an additional check, we found that \leff shifted by $<~0.005$ if the analysis threshold of $1$~keVee was lowered to 0.5~keVee.
\begin{table}[h]
\centering
\caption{\leff and the statistical uncertainty are tabulated for each spline point shown in Fig. \ref{fig3}.  Also shown are the systematic errors arising from uncertainty in the $S1$ peak-finding efficiency (``S1 eff.'') and the Xe(n,n)Xe elastic scattering cross-section (``$\sigma_{el}$~data''). These systematic uncertainties are discussed in Sec. \ref{S1_systematics} and Sec. \ref{MC_systematics}.}
\begin{tabular}{ccccc}
\\
\hline
~keVr~~&~ \leff~ & \multicolumn{3}{c}{{\it Uncertainty}}  \\
 & &~~statistical &~~$S1$~eff.~& $\sigma_{el}$~data \\
\hline
2 & 0.160 & $\pm0.014$ & $\pm0.012$ & $\pm0.009$ \\ 
5 & 0.156 &  $\pm0.011$ &  $\pm0.006$ & $\pm0.001$ \\ 
10 & 0.162 &  $\pm0.012$ &  $\pm0.005$ & $\pm0.002$ \\ 
15 & 0.194 & $\pm0.011$ &  $\pm0.006$ & $\pm0.005$\\ 
25 & 0.220 & $\pm0.012$ &  $\pm0.005$ & $\pm0.001$\\ 
50 & 0.237 & $\pm0.009$ &  $\pm0.004$ & $\pm0.001$\\ 
100 & 0.274 & $\pm0.010$ &  $\pm0.004$ & $\pm0.005$\\ 
\hline
\end{tabular}
\label{systm}
\end{table}
\subsection{Input to the Simulation} \label{MC_systematics}
The Monte Carlo simulation was performed using GEANT4.9.0.  By default this software uses the ENDF/B-VI nuclear cross section data, 
in which xenon cross sections are based on \cite{1967_Hodgson}.  
Instead, we used the updated ENDF/B-VII xenon nuclear cross section data to obtain the nuclear recoil spectrum, which is based on more recent calculations using the Optical Model Potential (OMP) parameters in \cite{ENDF-VII,key-8}.  
The uncertainty in our Monte Carlo recoil energy spectrum arising from uncertainty in the $\sigma_{el}$ data for Xe(n,n)Xe scattering was therefore determined by the OMP theory.
  
A comprehensive study of calculated versus measured differential neutron cross sections of nuclei from $A=19$ to $A=209$ \cite{2003_Konig} allowed the refinement of the OMP theory for nuclei in that mass range.  A related study \cite{key-3} concluded that for nuclei where experimental elastic scattering data is lacking, a $\pm3\%$ uncertainty in the OMP potential well depth parameter could be used to determine a conservative estimate of the uncertainty in the elastic cross sections.
We used the EMPIRE software package \cite{key-6} along with the best OMP model for xenon to calculate
$\mbox{d}^2{\sigma_{el}}/\mbox{d}E \mbox{d}\Omega$
for all relevant neutron energies.  The cases $-3\%$ and $+3\%$ in the OMP potential well depth parameter were considered, and new nuclear recoil energy spectra were obtained.  An average change of $<2.5\%$ in the nuclear recoil spectra was found, in the  energy range $1-100$~keVr.  

The \leff analysis was repeated for the two new spectra. The resulting measured energy dependence of \leff is shown in Fig. \ref{fig3} (solid red lines) along with the result using the default Monte Carlo energy spectrum.  This systematic uncertainty in \leff due to the input elastic scattering cross section data was found to be smaller than the statistical uncertainty.  Additionally, the older ENDF/B-VI cross section data resulted in an energy dependence of \leff that is  consistent with the result shown in Fig. \ref{fig3} for $2<E_{nr}<50$~keVr, and about $2\sigma$ lower at 100~keVr. 

We also note that the only available experimental measurements of the total
cross section for neutrons on xenon are from \cite{1960_Vaughn},
which are consistent with the OMP theory used in this paper. However, this experimental data provides weak overall constraints on the
uncertainties in the double differential elastic cross section. 

\subsection{$S1$ resolution} \label{S1_res_systematics}
The $S1$ resolution for nuclear recoils is dominated by binomial fluctuations in the collection of scintillation photons.  It is further impacted by the intrinsic single photo-electron resolution of the PMTs, which had an average $\sigma/\mu=0.58\pm0.05$.  From this, the $S1$ resolution was found to be $\times1.16$ larger than that expected solely from the binomial fluctuations, in the range $E_{nr}~<~300$~keVr \cite{2008_Dahl}.  This result was determined from the width of the log$_{10}(S2/S1)$ band.  The total light detection efficiency for $S1$ was measured to be $8-11$\%.   Considering the uncertainty in the PMT single photo-electron response, and taking a conservative estimate of $\pm20\%$ uncertainty in the $S1$ light detection, the uncertainty in the $S1$ resolution was calculated to be $<\pm2\%$ \cite{2008_Dahl}.  If the resolution were better (worse) by this amount, the \leff curve in Fig. \ref{fig3} would translate higher (lower) by less than 1/4 the statistical error.  This effect was included in the statistical uncertainty.

\subsection{Behavior of \leff}
The expectation that \leff should continue to fall with decreasing energy is due to the Lindhard model \cite{1963_Lindhard} for the partitioning of recoil energy into atomic motion and electronic excitation (so-called nuclear quenching).  In Ge the agreement between experiment and the Lindhard prediction is very good, above 10~keVr \cite{2007_Benoit}.  Below 10~keVr measured values are as much as $\times2$ above the Lindhard prediction \cite{1995_Messous}.  

\begin{figure}[h]
\includegraphics[width=0.48\textwidth]{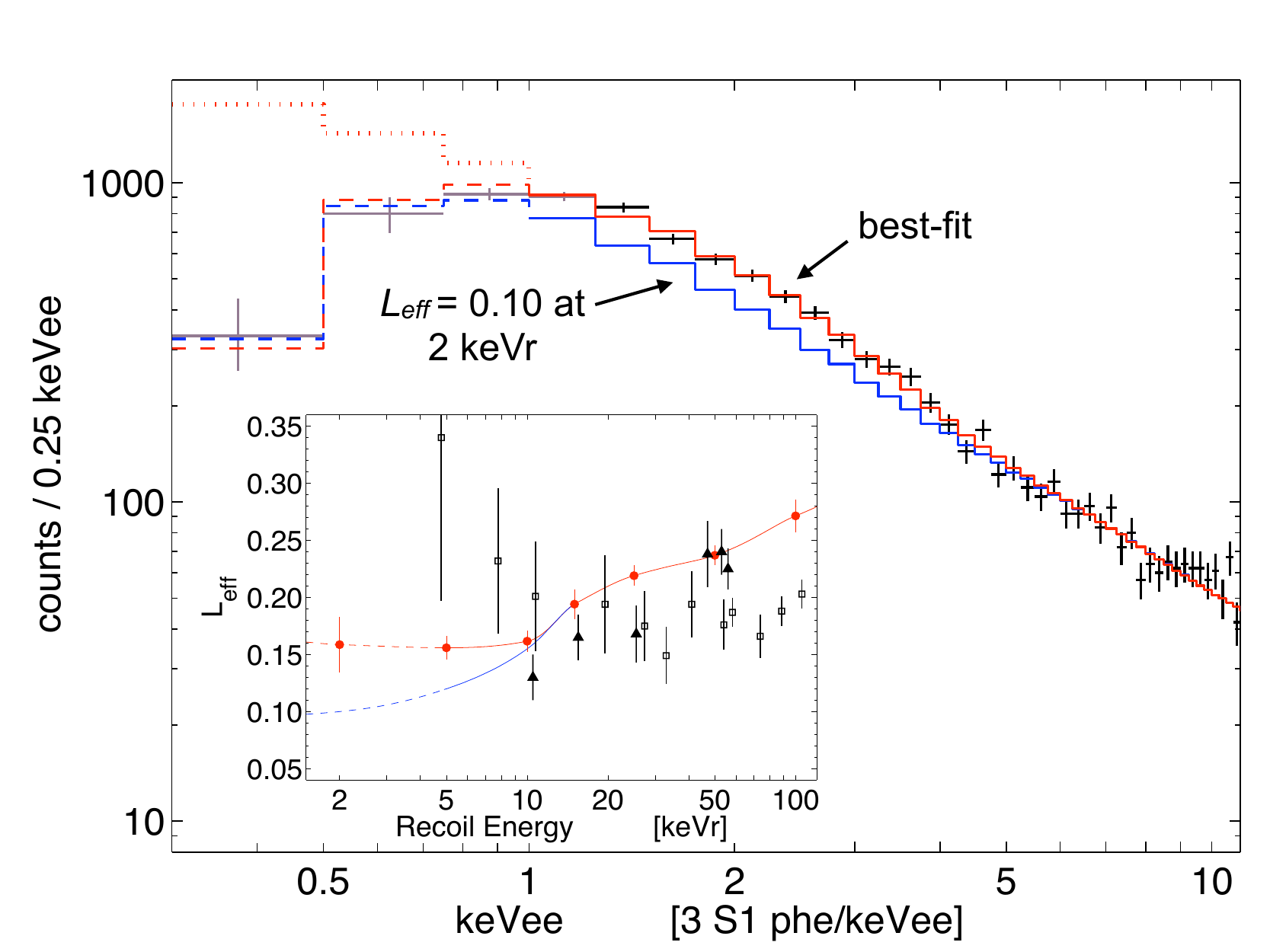}
\caption{A scenario in which \leff continues to fall below 10~keVr, to 0.10 at 2~keVr (as shown inset, blue line) is inconsistent with our data.  The best-fit Monte Carlo (red line) and energy dependence of \leff (inset, red line) are as shown in Fig. \ref{fig2} and Fig. \ref{fig3}.}
\label{fig7}
\end{figure}

In Fig. \ref{fig7} we show the shape of the simulated single scatter nuclear recoil spectrum if \leff were to continue to drop monotonically below 10~keVr.  This was done by taking the best-fit \leff shown in Fig. \ref{fig3}, and forcing the spline point at 2~keVr to have \leff$=0.10$.  The prediction of about 20\% fewer events per bin for all bins $1-3$~keVee is more than $5\sigma$ inconsistent with the data ($\chi^2>700$), showing that a decreasing \leff below 10 keVr is strongly disfavored by our data.   

The Lindhard model applies to the total detectable energy transferred to electronic excitation.  The Lindhard prediction of \leff in xenon is systematically about 0.03 higher than our result in the energy range $10-100$~keVr (see for example \cite{2007_Mangiarotti}).  This is not surprising since scintillation light is generated by electrons which recombine.  A theoretical model of biexcitonic electronic quenching has been proposed \cite{2005_Hitachi}.  This prediction for the total quenching of xenon scintillation is lower than the Lindhard prediction (when applied directly to scintillation), and is shown in Fig. \ref{fig3} (dash-dot line).  

It has also been noted \cite{2007_Mangiarotti, 2007_Hitachi} that for nuclear recoils below $\sim10$~keVr in xenon, the underlying Thomas-Fermi interaction potential may be a poor approximation, and the Lindhard model for nuclear quenching may not be valid.  This is roughly where our \leff result ceases to fall with decreasing energy, and may indicate the point at which the electronic stopping becomes comparable to the nuclear stopping \cite{2007_Hitachi}.

\section{Ionization Yield} \label{e_yield}
\begin{figure}[h]
\includegraphics[width=0.46\textwidth]{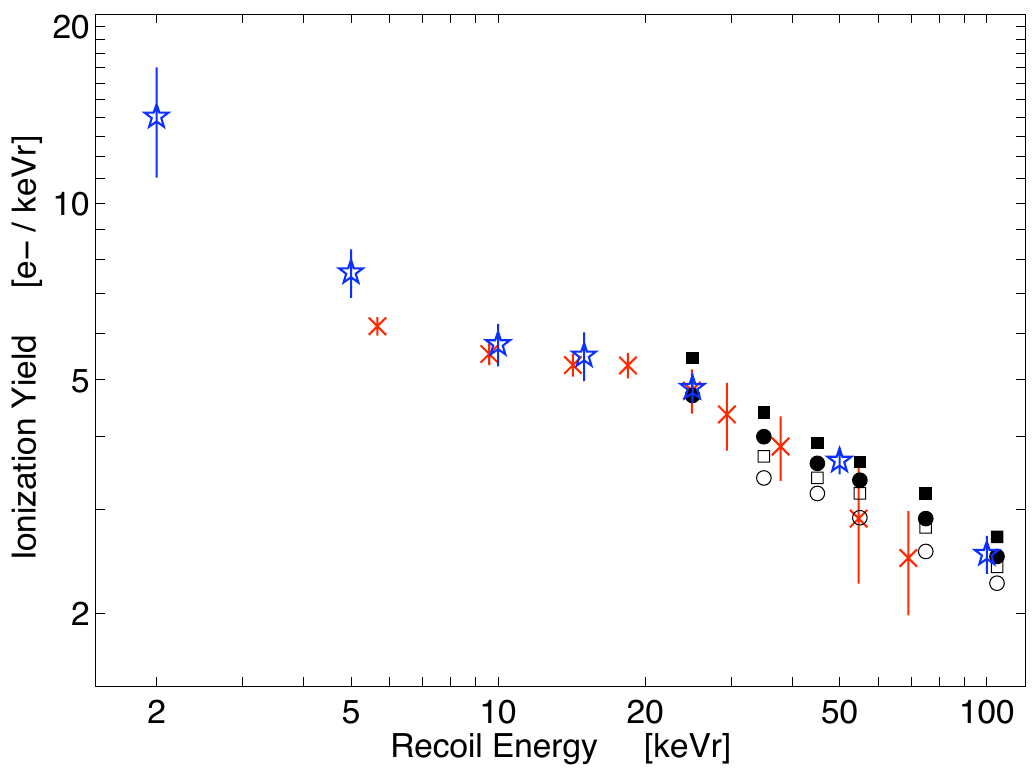}
\caption{The ionization yield \Qy (blue stars) of xenon for nuclear recoils at $E_d~=~0.73$~kV/cm, as determined from the energy dependence of \leff (shown in Fig. \ref{fig3}).  \Qy as determined from the multiple scatter method described in Sec. \ref{e_yield} is also shown (red $\times$), along with data from \cite{aprile2006} (filled and open circles/squares, uncertainty not shown).}
\label{fig8}
\end{figure}

The ionization yield \Qy is uniquely determined from the $S1$ and $S2$ data, once \leff is known.  This is shown in Fig. \ref{fig8} (blue stars) for each of the \leff spline points from Table \ref{systm}.  This result, for $E_d=0.73$~kV/cm, agrees with previous measurements \cite{aprile2006} above 20~keVr.  Data from \cite{aprile2006} are shown in Fig. \ref{fig8} as filled and open circles/squares, with the uncertainty omitted for clarity.  The filled and open data points indicate measurements from 2 different detectors, while the circles (squares) represent measurements at $E_d=0.10$ ($E_d=2.0$)~kV/cm.  The modest variation of \Qy with applied electric field is expected since \leff (and by inference $\mathcal{Q}_y$) has been measured \cite{aprile2005} to have only a weak ($<10\%$) dependence on the applied electric field.  
\subsection{Multiple Scatter Method} \label{msmethod}
A single ionization electron extracted to the gas was found to correspond to an $S2$ signal of $24\pm7$~photo-electrons \cite{sorensen2006}.  $S2$ pulses were clearly identifiable over $>5$~orders of magnitude in size, and as many as 4 scatters in the fiducial target were observed for some elastic nuclear recoil events.

The multiplicity ($m=1...3$) of an elastic nuclear recoil event is defined here as the number of scatters with recoil energy above a given threshold.  The analysis in Sec. \ref{scint_yield} was restricted to events with $m=1$ and an $S2$ threshold of 200~photo-electrons, or 8~ionization electrons.  The measured multiplicity is a sensitive function of the specified recoil energy threshold per scatter.  A specific $S2$ threshold ($S2_{thr}$) in the data was correlated with a keVr equivalent threshold ($E_{thr}$) from the Monte Carlo, by comparing the predicted and experimentally measured event multiplicities.  Due to the typical $0.8~\mu$s FWHM of $S2$ pulses, scatters with $dz<3$~mm were not easily resolvable, regardless of their separation in $(x,y)$.  This criteria was duplicated in the Monte Carlo analysis.
\begin{figure}[h]
\includegraphics[width=0.49\textwidth]{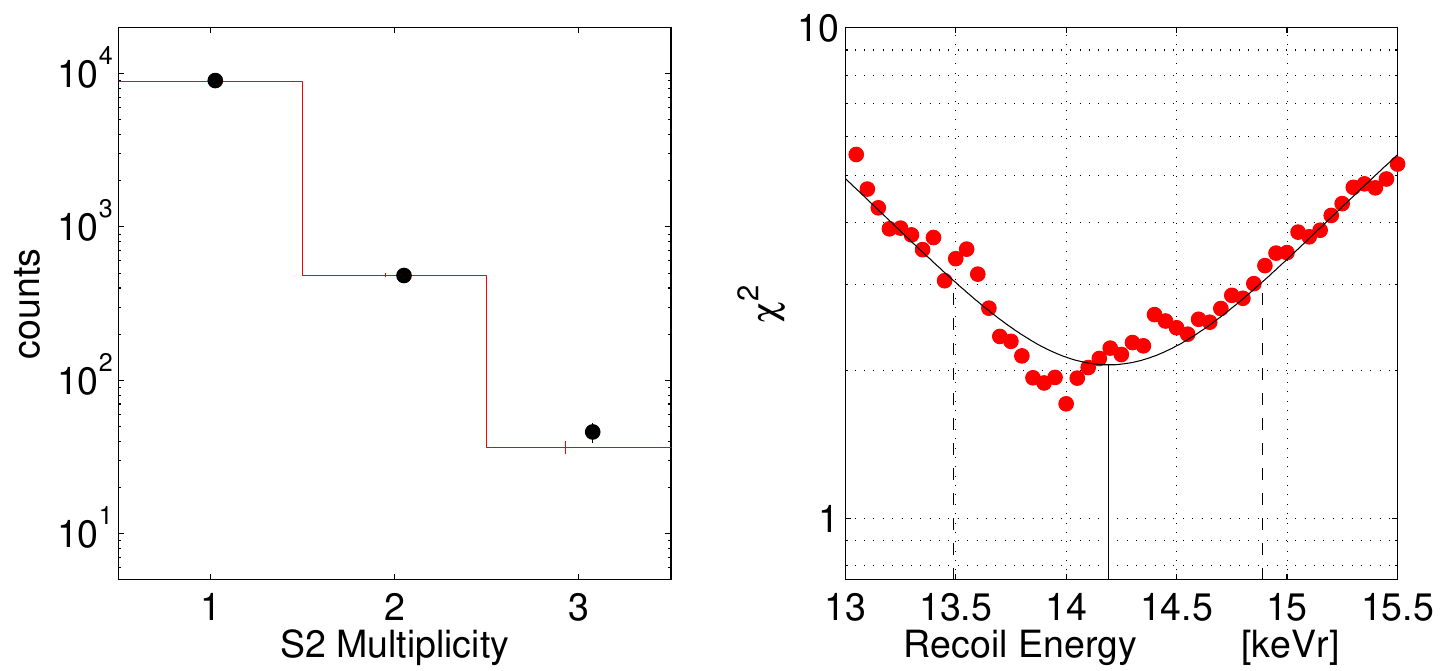}
\caption{ {\bf (left)} The number of single, double, etc. scatters in the fiducial target with $S2$ (per vertex) $>1600$~photo-electrons (black circles) is compared with the Monte Carlo prediction (red line).  Only scatters with an energy deposition per vertex $E>E_{thr}$ were counted.  The comparison is a sensitive function of the Monte Carlo energy threshold.  The best fit corresponds to $E_{thr}=14.2\pm0.7$~keVr.  {\bf (right)} The $\chi^2$ distribution (red circles) for the comparison shown at left as a function of the recoil energy threshold imposed in the Monte Carlo.  A $2^{nd}$-order polynomial fit (black line) was used to extract the minimum and the $1\sigma$ uncertainty (indicated by dashed lines).}
\label{fig9}
\end{figure}

An example of the multiplicity comparison is shown in Fig. \ref{fig9} (left) for $S2_{thr}=1600$ photo-electrons.  In each case, only events with all scatters in the 5.4~kg fiducial target were counted.  The best fit between the number of scatters for $m=1...3$ in the data and the Monte Carlo was obtained for $E_{thr}=14.2\pm0.7$~keVr.  The overall normalization and the recoil energy threshold were free parameters, so the minimum $\chi^2=2$ with $d.o.f.=1$ gives a $p$-value of 0.16.  The $\chi^2$ distribution for the data versus Monte Carlo comparison of scatter multiplicity is shown as a function of $E_{thr}$ in Fig. \ref{fig9} (right).

The example shown in Fig. \ref{fig9} corresponds to the data point (red $\times$) at 14.2~keVr in Fig. \ref{fig8}.  The ionization yield was calculated as \Qy$=1600/24/14.2=4.7$~ionization electrons/keVr.  This value was then corrected to account for the measured $S2$ resolution, since $E_{thr}$ from the Monte Carlo has perfect resolution.  The correction increases \Qy by an average of 11\%.  The error bars include the uncertainty in the mean of the single-electron $S2$ peak, and the uncertainty from each fit (as shown in Fig. \ref{fig9}, right).  The $\chi^2$ distributions for all data points shown in Fig. \ref{fig8} exhibited a clear minimum, a $\chi^2/d.o.f.\sim1$ and were well-fit by a $2^{nd}-$order polynomial.  

\subsection{Discussion}
The major benefit of the multiple scatter method described in Sec. \ref{e_yield} is that it provides an independent check on the \leff result shown in Fig. \ref{fig3}.  Above 5~keVr, the multiple scatter method (Fig. \ref{fig8}, red $\times$) predicts a \Qy that is consistent with the values inferred from the best-fit $\mathcal{L}_{eff}$.  Below $\sim5$~keVr, the multiple scatter method predicts a constant \Qy$\sim6$~ionization electrons/keVr (not shown in Fig. \ref{fig8}).  In this regime, the result of the method is affected by the light collection for primary scintillation ($S1$).  This somewhat unexpected result can be understood from Fig. \ref{fig4}.  Lines of constant $S2$ lie parallel to the line corresponding to $S2=200$~photo-electrons (8 ionization electrons), which is shown dashed in the lower left.  For example, at $E_{thr}=2$~keVr, $S2_{thr}\approx300$~photo-electrons and a substantial number of events used in the comparison have $E<1$~keVee.  This implies that an additional number of events were lost (due to the $S1$ signal acceptance).  The multiplicity of these events is not known, and this likely affects the determination of $\mathcal{Q}_y$.

\section{Summary} \label{summary_sec}
The energy threshold of a liquid xenon time-projection chamber using proportional scintillation for $S2$ is presently determined by its sensitivity to primary scintillation ($S1$) and by $\mathcal{L}_{eff}$.  We have presented the energy dependence of \leff based on an analysis of nuclear recoils in the XENON10 experiment.  XENON10 recently reported a WIMP-nucleon (elastic scattering) exclusion limit of $\sigma=4.5\times10^{-44}~(8.8\times10^{-44})$~cm$^2$ for a WIMP mass of 30~(100)~GeV/c$^2$ \cite{2008XENON10SI} assuming a constant \leff$=0.19$.  A conservative upper bound of $\sigma=5.2\times10^{-44}~(10.4\times10^{-44})$~cm$^2$ was quoted in \cite{2008XENON10SI} based on a preliminary version of this work.  Considering the energy dependence of \leff shown in Fig. \ref{fig3}, the XENON10 WIMP-nucleon exclusion limit was found to be $\sigma=5.1\times10^{-44}~(10.2\times10^{-44})$~cm$^2$ for a WIMP mass of 30~(100)~GeV/c$^2$.

We also presented the absolute ionization yield of xenon for nuclear recoils ($\mathcal{Q}_y$), as a function of recoil energy.  For XENON10, this clearly shows that the trigger threshold for nuclear recoils is as low as $\sim1$~keVr.  The analysis threshold is presently limited by the light collection for primary scintillation ($S1$).  The technique used to determine \leff for primary scintillation ($S1$) should also be applicable to the secondary scintillation ($S2$);  a future work will explore this possibility. 

\appendix

\section*{Acknowledgments}
We are thankful to LNGS for their support during the commissioning and operation of XENON10, and Yale University Biomedical High Performance Computing Center where the simulations were performed.  This work was funded by NSF Grants No. PHY-03-02646 and No. PHY-04-00596, the CAREER Grant No. PHY-0542066, the DOE Grant No. DE-FG02-91ER40688, NIH Grant No. RR19895, by the Swiss National Foundation SNF Grant No. 20-118119, by the Volkswagen Foundation (Germany) and by the FCT Grant No. POCI/FIS/60534/2004 (Portugal).

\end{document}